\documentclass[10pt,a4paper]{article}
\usepackage[ansinew]{inputenc}
\usepackage[hmargin=2cm, vmargin=2cm]{geometry}
\usepackage [T1]{fontenc} 
\usepackage{textcomp}
\usepackage{amsmath,amsfonts,latexsym, amssymb,amsthm}
\usepackage{array} 
\usepackage{hyperref}
\usepackage[sort&compress]{natbib}
\usepackage{stmaryrd} 

\bibpunct{[}{]}{,}{n}{}{;} 

\newcommand{\Prob}{{\rm P}}

\newcommand{\ident}{\stackrel{\mbox{{\rm \tiny (def)}}}{=}}
\newcolumntype{C}{>{$}c<{$}}
\newcolumntype{L}{>{$}l<{$}}
\newcolumntype{R}{>{$}r<{$}}

\newtheorem* {theorem*}{Theorem}
\newtheorem* {corollary*}{Corollary}
\newtheorem* {definition*}{Definition}

\title{\bf
Polynomial time factoring algorithm\\
using Bayesian arithmetic
}
\author {Michel Feldmann
\thanks{Electronic address: michel.feldmann@polytechnique.org}
}
\date {}

\begin{document}

\maketitle
\abstract{
In a previous paper, we have shown that any Boolean formula can be encoded as a linear programming problem in the framework of Bayesian probability theory. When applied to \textsf{NP}-complete algorithms, this  leads to the fundamental conclusion that \textsf{P} = \textsf{NP}. Now, we implement this concept in elementary arithmetic and especially in multiplication. This provides a polynomial time deterministic factoring algorithm, while no such algorithm is known to day. This result clearly appeals for a revaluation of the current cryptosystems. The Bayesian arithmetic environment can also be regarded as a toy model for quantum mechanics.
} 


\section{Introduction}
\label{introduction}
Arithmetic  is a part of abstract mathematics, i.e., a theory based, for instance, on Peano axioms and dealing with infinitely many elements. On the other hand, arithmetic is also a practical way of counting and computing. Paradoxically,  computation rules are not very much concerned by the abstract theory. Instead, only Boolean operations are at work, e.g., in micro-processors. In a previous paper~\cite{mf3}, we have shown that Boolean operations can be described in terms of Bayesian probability leading to the stunning result \textsf{P} = \textsf{NP}. We propose in the present paper to investigate the Bayesian structure of a particular environment, namely, elementary arithmetic operations. 

In this approach,  we start with a Boolean algebra composed of all relevant binary digits involved in a Diophantine equation. In order to deal with tractable formulae, it is suitable to introduce internal variables beforehand, e.g. carry bits. Now, we construct a Bayesian environment and we account for the rules of arithmetic by means of structural equations. These constraints will be further added to a number of specific equations corresponding to the input data  and a number of universal equations reflecting the laws of logics. The Bayesian method consists in checking the consistency of all these conditions  by linear programming (LP). When feasible, the undefined bits are eventually computed. 

The model is founded on the theory of probability. Nevertheless, when the LP problem is feasible in the \emph{general environment},  the system always accept strictly deterministic solutions. This is easily proved~\cite{mf3} by exploring the full ensemble of possible assignments.  By contrast, in the present \emph{arithmetic environment}, the use of internal variables impose a limitation in the set of accessible assignments, because internal variables cannot be assigned independently. Therefore only a part of the potential assignments can be consistently explored and the proof is no longer valid.  As a result, a feasible LP problem may or not accept deterministic solutions.  For instance, in the factoring algorithm, the LP problem is generally feasible: When the input integer is composite, its factors are derived from the deterministic solutions. On the contrary, when the input is prime, we have no deterministic non-trivial factors but we do have probabilistic solutions. In this respect, the Bayesian system can be regarded as a toy model of quantum formalism: a `composite system' is likened to a classical object with deterministic parameters while a `prime system' is likened to a quantum object with only probabilistic parameters. We will shortly sketch an example in Sec.~\ref{factorization}. Fortunately, LP is quite efficient to compute the existing deterministic solutions or decide with certainty that no such solution exists. Thus, we obtain both a deterministic polynomial time factoring algorithm and a deterministic polynomial time criterion of primality. Presently, only a quantum algorithm is known for the first case~\cite{shor} and a class-\textsf{P} algorithm was only found recently~\cite{agrawal}  for the second case.

The general framework of the theory is the following: We consider a Boolean binary algebra with $N$ variables $\mathsf{X}_i$, for $i\in \llbracket 1,N\rrbracket$.  Thus, we may potentially assign a value $0$ or $1$ to each variable. We name \emph{complete assignment} a full assignment to the $N$ variables and \emph{partial assignment} an assignment to less than $N$ variables. We note   $\overline{\mathsf{X}}_i$ the negation of $\mathsf{X}_i$,  and call \emph{literal} a variable or its negation. Given two logical formulae (or \emph{decision functions}) $\mathsf{f}_1$ and $\mathsf{f}_2$, it is convenient to note  $(\mathsf{f}_1 ; \mathsf{f}_2)$ (with a semicolon) the conjunction $\mathsf{f}_1\wedge\mathsf{f}_2$ and  $(\mathsf{f}_1 , \mathsf{f}_2)$ (with a comma) the disjunction $\mathsf{f}_1\vee\mathsf{f}_2$. We name \emph{requirement} a conjunction of literals, \emph{complete requirement} a conjunction of $N$ literals, e.g., $\Xi=({\mathsf{X}}_1;\overline{\mathsf{X}}_2;\dots;{\mathsf{X}}_N)$, that is satisfiable by a complete assignment, e.g., $\xi=(1, 0,\dots, 1)$, and \emph{partial requirement} a conjunction of less than $N$ literals, e.g., $({\mathsf{X}}_i;\overline{\mathsf{X}}_j;{\mathsf{X}}_k)$. Clearly, there are $2^N$ different complete assignments and therefore $2^N$  complete requirements. On the other hand, with up to $N$ variables, it is possible to construct $2^{2^N}$ different decision functions.

Now, we propose to regard any \emph{decision function} as a \emph{random event} and to reformulate the logical equations as a set of linear equations between the \emph{probabilities} of the relevant requirements. For this, we use the Bayesian conception of the theory of probability~\cite{jaynes}. Given by hypothesis that a particular logical proposition $({\Lambda})$ has to be satisfied, \emph{the probability of any event will be conditioned by $({\Lambda})$}. For instance, in the conventional addition of two integers $U$ and $V$, $({\Lambda})$ will be the statement $({\Sigma})$ that the two integers $U$ and $V$ sum to a third integer $S$.

The basic probability set is the ensemble $\Omega = \{ \Xi\}$ of all  $2^N$ complete requirements, labelled by the $2^N$ complete assignments $\xi$. Since the cardinality of $\Omega$ is finite, the power set $\mathcal{P}(\Omega)$, of cardinality $2^{2^N}$, is a sigma-algebra $\mathcal{T}$, identical to  the ensemble of all decision functions. Now we have to define a probability distribution $\Prob$ on $\mathcal{T}$ conditioned by $({\Lambda})$. Finally, the Kolmogorov probability space is $(\Omega, \mathcal{T}, \Prob)$.

We start with the prior information that $({\Lambda})$ is TRUE and determine how this knowledge affects the conditional probability of the relevant requirements. It turns out that these constraints are conveniently formulated as a LP problem. Therefore, we complete the computation by solving this LP problem. For \textsf{NP}-algorithms,  the number of relevant requirements scales as O$(N^K)$, where $K$ is an integer. Thus, the LP solutions are obtained in polynomial time.

In this paper, we will investigate the behaviour of elementary arithmetic operations in such a Bayesian context. We will first implement with full details the \emph{Bayesian addition}. Needless to say that the method is completely maladjusted for practical operations and cannot compete with a direct computation.  However the derivation is quite simple and adequate to clarify the present concept. Furthermore, Bayesian addition is a part of \emph{Bayesian multiplication}: Again, the method is by far too complicated when compared with a direct product but the interest lies in the inverse problem, namely factorization. Even for this last problem, the method, at least in its present state, is complicated for small numbers. But the unique feature is clearly the scaling capability: Factorization of an integer of $n$ bits is obtained by LP in a system of O$(n^2)$ unknowns. This remains the only possibility to factorize integers of hundreds or even thousands of bits.


\section{Addition of two integers}

Let $U$ and $V$ be two integers. Without loss of generality we can assume that they are both described by the same number $n$ of bits, given that we may complete by a number of zeros if necessary. Let $S=U+V$ be the sum. The binary expansions read,

$$ {U}= \sum_{i=0}^{n-1} {u}_i 2^i \ ; \ {V}= \sum_{i=0}^{n-1} {v}_i 2^i \  ; \ {S} = {U} + {V}= \sum_{i=0}^{n} {s}_i 2^i $$
with $u_i, v_i, s_i\in \{ 0,1\}$. It is suitable to introduce carry bits explicitly.

\emph{Example:} Let $n=2$ and let $r_i$ be the carry bits. The binary operation can be written  as 
\begin{center}
\begin{tabular}{L|CCC}
U                            &              & {u_{1}} &   {u_{0}} \\
V                            &    & {v_{1}} & {v_{0}}\\
R                   &  {r_{2}}   & {r_{1}}  &  . \\
\hline
S &  {s_{2}}   & {s_{1}} & s_{0} \\
\end{tabular}
\end{center}
In Bayesian arithmetic, all bits, including the carry bits are considered as random variables. The assignments $u_i$, $v_i$, $s_i$ and $r_i$ are likened to the outcomes of these random variables, respectively $\mathsf{U}_i$, $\mathsf{V}_i$, $\mathsf{S}_i$ and $\mathsf{R}_i$.

 Let $\Sigma$ be the logical proposition: `${S}$ is the sum of ${U}$ and ${V}$'.  \emph{We will compute the conditional probabilities of all events given $\Sigma$}. 

We have first to define the input data. Provisionally, we suppose that $U$ and $V$ are given, but it would be possible to choose different inputs, e.g., $S$ and $U$ or even exotic data like carry bits.
Then, for $i\in\llbracket 0,n-1\rrbracket$ we suppose that $\mathsf{U}_i =u_i$ and $\mathsf{V}_i =v_i$ with certainty, i.e., with a probability $1$. The probabilistic formulation is

\begin{align}
\Prob(\mathsf{U}_i= u_i|\Sigma)&=1    \\
\Prob(\mathsf{V}_i= v_i|\Sigma)&=1.    
\end{align}
The partial probabilities, e.g., $\Prob(\mathsf{U}_i= u_i|\Sigma)$ are regarded as the \emph{unknowns} of the problem. More generally, we will call \emph{partial probability} the probability of any requirement, conditioned in this section by $\Sigma$, and identify such partial probabilities with unknowns (not to be mistaken for the very \emph{variables}, like e.g. $\mathsf{U}_i$).
This codification define a set of $2n$ linear equations that we will call \emph{data specific equations}. We will next define \emph{structural equations}, expressing the rules of arithmetic, and later \emph{universal equations}, expressing  the laws of logics.

In order to construct the structural equations, let us consider the following \emph{one-bit full adder  truth table}.                            

\begin{center}
\begin{tabular}{||C|C|C||C|C||}
\hline
\hline
\mathsf{U}_i&\mathsf{V}_i&\mathsf{R}_i&\mathsf{S}_i&\mathsf{R}_{i+1}\\
\hline
0&0&0&0&0\\
0&0&1&1&0\\
0&1&0&1&0\\
0&1&1&0&1\\
1&0&0&1&0\\
1&0&1&0&1\\
1&1&0&0&1\\
1&1&1&1&1\\
\hline
\hline
\end{tabular}
\end{center}
For $i=0$, the outcome of the variable $\mathsf{R}_0$ is always zero with certainty and therefore this variable will be omitted. For $i=n$ we have
\begin{equation}
\label{sn}
\Prob(\mathsf{S}_{n}=1|\Sigma) = \Prob(\mathsf{R}_{n}=1|\Sigma)
\end{equation}
Therefore, we will not discriminate between $\mathsf{S}_{n}$ and $\mathsf{R}_{n}$. It will be convenient to keep rather $\mathsf{R}_{n}$ and simply omit the variable $\mathsf{S}_{n}$. 

\emph{Example: } 
Let $n=2$. We will deal with the following variables:
 $\mathsf{U}_0$, $\mathsf{U}_1$, $\mathsf{V}_0$, $\mathsf{V}_1$, $\mathsf{S}_0$, $\mathsf{S}_1$, $\mathsf{R}_1$, $\mathsf{R}_2$.
The truth tables for $i= 1, 0$ read respectively

\begin{center}
\begin{tabular}{||C|C|C||C|C||}
\hline
\hline
\mathsf{U}_1&\mathsf{V}_1&\mathsf{R}_1&\mathsf{S}_1&\mathsf{R}_{2}\\
\hline
0&0&0&0&0\\
0&0&1&1&0\\
0&1&0&1&0\\
0&1&1&0&1\\
1&0&0&1&0\\
1&0&1&0&1\\
1&1&0&0&1\\
1&1&1&1&1\\
\hline
\hline
\end{tabular}
\begin{tabular}{||C|C||C|C||}
\hline
\hline
\mathsf{U}_0&\mathsf{V}_0&\mathsf{S}_0&\mathsf{R}_{1}\\
\hline
0&0&0&0\\
0&1&1&0\\
1&0&1&0\\
1&1&0&1\\
\hline
\hline
\end{tabular}
\end{center}

Coming back to the general case, we undertake to translate the truth tables into linear equations between partial probabilities. This is straightforward because the probability of an union of mutually exclusive events is the sum of the probability of each event. 

For instance, we read that $(\mathsf{S}_{i}=1)$ is the union of four mutually exclusive conjunctions, namely, $[(\mathsf{U}_{i}=1)\wedge(\mathsf{V}_{i}=1)\wedge(\mathsf{R}_{i}=1)]$, $[(\mathsf{U}_{i}=1)\wedge(\mathsf{V}_{i}=0)\wedge(\mathsf{R}_{i}=0)]$, $[(\mathsf{U}_{i}=0)\wedge(\mathsf{V}_{i}=1)\wedge(\mathsf{R}_{i}=0)]$ and $[(\mathsf{U}_{i}=0)\wedge(\mathsf{V}_{i}=0)\wedge(\mathsf{R}_{i}=1)]$. Therefore, the probability of $(\mathsf{S}_{i}=1)$ is the sum of the probabilities of the four conjunctions. 
Note that we are not concerned by the event $(\mathsf{S}_{i}=0)$ because its probability is \emph{logically} connected with the probability of $(\mathsf{S}_{i}=1)$. This will be accounted for later, by 
the universal equations.

Now we can construct the structural equations of the addition environment by inspection of the truth tables.

- For $i \in\llbracket 1, n-1\rrbracket$ we obtain $2(n-1)$ equations,

\begin{align*}
\begin{aligned}
\Prob(\mathsf{S}_i=1|\Sigma)&= \Prob( \mathsf{U}_i =0;\mathsf{V}_i=0;\mathsf{R}_i=1|\Sigma) +\Prob( \mathsf{U}_i=0;\mathsf{V}_i=1;\mathsf{R}_i=0|\Sigma) \\
&+ \Prob( \mathsf{U}_i=1;\mathsf{V}_i=0;\mathsf{R}_i=0|\Sigma) + \Prob( \mathsf{U}_i=1;\mathsf{V}_i=1;\mathsf{R}_i=1|\Sigma) 
\end{aligned}
\\
\begin{aligned}
 \Prob(\mathsf{R}_{i+1}=1|\Sigma)&= \Prob( \mathsf{U}_i =0;\mathsf{V}_i=1;\mathsf{R}_i=1|\Sigma) +\Prob( \mathsf{U}_i=1;\mathsf{V}_i=0;\mathsf{R}_i=1|\Sigma) \\
&~~+ \Prob( \mathsf{U}_i=1;\mathsf{V}_i=1;\mathsf{R}_i=0|\Sigma) + \Prob( \mathsf{U}_i=1;\mathsf{V}_i=1;\mathsf{R}_i=1|\Sigma)
\end{aligned}
\end{align*}

- For $i=0$, we have two particular equations: Since $\mathsf{R}_0$ is omitted, we have only one or two relevant conjunctions in each equation,
\begin{align*}
 \Prob(\mathsf{S}_0=1|\Sigma)&= \Prob( \mathsf{U}_0=0;\mathsf{V}_0=1|\Sigma) + \Prob( \mathsf{U}_0=1;\mathsf{V}_0=0|\Sigma) \\
 \Prob(\mathsf{R}_{1}=1|\Sigma)&= \Prob( \mathsf{U}_0 =1;\mathsf{V}_0=1|\Sigma). 
\end{align*}

- For $i=n$, since we have omitted $\mathsf{S}_{n}$ Eq.(\ref{sn}) is unnecessary and we have nothing to set down.

We have then completed the construction of the $2n$ \emph{structural equations}. With the $2n$ specific data equations we have a total of $4n$ equations. 

On the other hand, we have $4n$ random \emph{variables}, namely $\mathsf{U}_i$ for $i \in\llbracket 0, n-1\rrbracket$, $\mathsf{V}_i$ for $i \in\llbracket 0, n-1\rrbracket$, $\mathsf{S}_i$ for $i \in\llbracket 0, n-1\rrbracket$ and  $\mathsf{R}_i$ pour $i \in\llbracket 1, n\rrbracket$. It is suitable to bring together these $4n$ variables in a single list $\mathsf{X}_k$ of \emph{global variables}, labelled from $k=1$ to $k=4n$. We will adopt throughout the labelling convention of table~\ref{varglobsimple}. 
\begin{table}
{\footnotesize
\begin{center}
\begin{tabular}{||C|C|C|C|C||}
\hline
\hline
\mathsf{U}_i  &  \mathsf{V}_i &  \mathsf{S}_i & \mathsf{S}_n=\mathsf{R}_n & \mathsf{R}_i \\
\hline
i\in\llbracket 0,n-1\rrbracket & i\in\llbracket 0,n-1\rrbracket & i\in\llbracket 0,n-1\rrbracket & i=n& i\in\llbracket 1,n\rrbracket \\
\hline
 k=i+1& k=i+n+1& k=i+2n+1& k=4n& k=i+3n\\
\hline
k\in\llbracket 1,n\rrbracket & k\in\llbracket n+1,2n\rrbracket & k\in\llbracket 2n+1,3n\rrbracket & k=4n& k\in\llbracket 3n+1,4n\rrbracket \\
\hline
 \mathsf{X}_k &  \mathsf{X}_k  &  \mathsf{X}_k  & \mathsf{X}_{4n}& \mathsf{X}_k \\
\hline
\hline
\end{tabular}
\end{center}
\caption{\footnotesize
\label{varglobsimple}
Labelling convention of the global variables $\mathsf{X}_k$ corresponding to the variables $\mathsf{U}_i$, $\mathsf{V}_i$ , $\mathsf{S}_i$ and $\mathsf{R}_i$ in the addition of two integers $U$ et $V$}
}
\end{table}
For instance, we have
\begin{align*}
 \mathsf{X}_{k}\ident\mathsf{S}_{k-2n-1} \mathrm{~~if~~}k \in\llbracket 2n+1, 3n\rrbracket\\
\end{align*}

\emph{Example}: Let $n=2$. The addition can be written in terms of global variables $\mathsf{X}_k$ as,

\begin{center}
\begin{tabular}{L|CCC}
U                            &              & \mathsf{X}_{2} &   \mathsf{X}_{1} \\
V                            &    & \mathsf{X}_{4} & \mathsf{X}_{3}\\
R                   &  \mathsf{X}_{8}   & \mathsf{X}_{7}  &  . \\
\hline
S &  \mathsf{X}_{8}   & \mathsf{X}_{6} & \mathsf{X}_{5} \\
\end{tabular}
\end{center}
Thanks to this notation, it is possible to make use of a shortcut: We will simply write  ($ k$) to describe the formula ($\mathsf{X}_k= 1|   \Sigma$), omitting both the reference to $\mathsf{X}$ and  the condition $\Sigma$. Similarly, we will note ($ -k$) for ($\mathsf{X}_k= 0|  \Sigma$). Now, we replace 
 $(\mathsf{U}_i=1|  \Sigma)$ by $(i+1)$ and $(\mathsf{U}_i =0|  \Sigma)$ by $(-i-1)$. 
The formula  $(\mathsf{V}_i=1|  \Sigma)$ is replaced by $(i+1+n)$ , $(\mathsf{S}_i=1|  \Sigma)$ is replaced by  $(i+1+2n)$ and for $i>0$, $(\mathsf{R}_i=1|  \Sigma)$ is replaced by $(i+3n)$. 
Finally, the set of structural equations are gathered together in Table \ref{equasimple} (where $j=i+1$ for simplicity).

\begin{table}
\footnotesize{
\fbox{
\begin{minipage}{\textwidth}
\begin{align*}
 &\Prob(2n+1)= \Prob( -1;n+1) + \Prob(1;-n-1) \\
&\Prob(2n+j)= \Prob( -j;-j-n;j+3n-1) +\Prob(-j;j+n;-j-3n+1) + \Prob( j;-j-n;-j-3n+1) + \Prob(j;j+n;j+3n-1)\\
 &\Prob(3n+1)= \Prob(1;n+1) \\
&\Prob(3n+j)= \Prob(-j;j+n;j+3n-1) +\Prob(j;-j-n;j+3n-1)+ \Prob(j;j+n;-j-3n+1) + \Prob(j;j+n;j+3n-1) \\
\end{align*}
\end{minipage}
}
\caption{\footnotesize
\label{equasimple}
Structural equations of addition expressed in terms of global variables ($ j\in\llbracket 2,n\rrbracket$). We have a total of $2n$ structural equations. 
}
}
\end{table}
To sum up, we have $2n$ specific data equation to specify the two input integers $U$ and $V$ and $2n$ structural equations to describe the  arithmetic operation of addition. 

\emph{Example}: Let $n=2$ and consider the addition $2+3$. 
We have $u_0=0$, $u_1=1$, $v_0=1$ and $v_1=1$. The  $4n=8$ equations read:
\begin{align}
\begin{aligned}
\label{n2specificplus}
&\Prob(1)=0 ;\
\Prob(-2)=0 ;\
\Prob(-3)=0 ;\
\Prob(-4)=0. \\
&\Prob( 5 ) = \Prob( 1 ;-3 )  + \Prob(-1 ; 3 ) \\
&\Prob( 6 ) = \Prob(-2 ;-4 ; 7 ) + \Prob(-2 ; 4 ;-7 ) + \Prob( 2 ;-4 ;-7 ) + \Prob( 2 ; 4 ; 7 )\\
&\Prob( 7 ) = \Prob( 1 ; 3 )\\
&\Prob( 8 ) = \Prob(-2 ; 4 ; 7 ) + \Prob( 2 ;-4 ; 7 ) + \Prob( 2 ; 4 ;-7 ) + \Prob( 2 ; 4 ; 7 )\\
\end{aligned}
\end{align}
(We have arbitrary chosen to formulate the data equations with zero right hand side).

\emph{Remark:} We have  presented the conventional addition of two integers $U$ and $V$. Actually, the Bayesian addition also fits  onto operations where the input data are not necessarily the bits of $U$ and $V$ but any set of assignments among the $4n$ variables. Let $n_b$ be this number of data bits. We have then $2n+n_b$  equations,
\begin{center}
\begin{tabular}{||l|c||}
\hline
\hline
Equations & number  \\
\hline
data   & $n_b$                   \\
structural  & $2n$                  \\
Total& $2n+n_b$                  \\
\hline
\hline
\end{tabular}
\end{center}
For instance, if $n=2$ and if we consider the subtraction $5-2$, we have $n_b=5$ and the data specific equations read
\begin{align}
\label{n2dataminus}
&\Prob(1)=0 ;\
\Prob(-2)=0 ;\
\Prob(-5)=0 ;\
\Prob(6)=0 ; \
\Prob(-8)=0. \
\end{align}
More generally, we may even specify any set of relevant unknowns, that we are going to define.

\paragraph{Relevant unknowns:} 
The unknowns of the LP problem are primary the partial probabilities involved in the set of specific or structural equations. However, these relevant partial probabilities are also involved in universal equations expressing the rules of logics. For instance the equation $\Prob(k)=0$ entails the logical consequence $\Prob(-k) =1$. We will name $\Prob(-k) $ a \emph{variant} of $\Prob(k) $. Similarly, the use of the relevant unknown $\Prob(k_1;k_2) $ entails the need to account for the logical consequence  $\Prob(k_1) = \Prob(k_1;k_2)+\Prob(k_1;-k_2)$ and we will also name $\Prob(k_1) $ and $\Prob(k_1;-k_2)$ \emph{variants} of $\Prob(k_1;k_2)$. In practice, we list all partial probabilities involved in all specific and structural equations: {We obtain the variants by iteration in removing one or several literals or in switching a literal into its negation}.
 
\emph{The {relevant unknowns} are the partial probabilities involved in the specific or structural equations or the variants of these partial probabilities}. 

In order to list the relevant unknowns, let us define a \emph{positive unknown} as an unknown involving only variables and no negation. For instance,  $\Prob(k_1; k_2;\dots)$ will be called positive if  and only if $k_1, k_2,\dots >0$. Now, let us inspect the different equations:

- We have $4n$ variables $\mathsf{X}_k$ and thus $4n$ positive relevant unknowns $\Prob(k)$ with one literal and finally $8n$ variants with one literal. 

- For the bit $i=0$, we have introduced $\Prob( 1;n+1)$, i.e., one positive unknown and thus 4 variants with two literals.

- For each bit $i \in\llbracket 1, n-1\rrbracket$ we have introduced $\Prob(i+1;i+1+n;i+3n)$, i.e., one positive unknown of three literal, and thus 3 positive variants with two literals, and finally 8 variants of three literals and $3\times 4$ variants of 2 literals.

\begin{table}
\footnotesize{
\fbox{
\begin{minipage}{\textwidth}
\begin{align*}
&\Prob(k)\\
&\Prob(1 ;n+1)\ ; \
\Prob(i+1;i+1+n) \ ;\  \Prob(i+1;i+3n) \ ;\  \Prob(i+1+n;i+3n)\\
&\Prob(i+1;i+1+n;i+3n)
\end{align*}
\end{minipage}
}
\caption{\footnotesize
\label{posiunknownsadd}
Relevant positive unknowns involved in the addition of $2$ integers of $n$ bits.
($i \in\llbracket 1, n-1\rrbracket$ ;
$k\in\llbracket 1,4n\rrbracket$). We have a total of $8n-3$ positive unknowns.
}
}
\end{table}

Collecting these results in Table \ref{posiunknownsadd}, the numbers of relevant unknowns are the following:
\begin{center}
\begin{tabular}{||c|c|c||}
\hline
\hline
Literals& positive unknowns    & unknowns  \\
\hline
1& $4n$                    & $8n$ \\
2& $3n-2$                  & $12n-8$\\
3& $n-1$                  & $8n-8$ \\
Total& $8n-3$                 & $28n-16$ \\
\hline
\hline
\end{tabular}
\end{center}

\emph{Example}: Let $n=2$. We have $28n-16=40$ relevant unknowns, namely:
\begin{description}
\item $\Prob( \pm 1$, $\Prob( \pm 2)$,  $\Prob(\pm 3)$,  $\Prob(\pm 4)$,  
$\Prob(\pm 5)$, $\Prob(\pm 6)$, $\Prob(\pm 7)$, $\Prob(\pm 8)$. 
\item $i=0$: $\Prob( \pm 1;\pm 3)$, 
\item $i=1$: $ \Prob( \pm 2;\pm 4;\pm 7)$,  
 $\Prob( \pm 2;\pm 4)$, $\Prob( \pm 4;\pm 7)$, $\Prob( \pm 2;\pm 7)$. 
\end{description}
The $8n-3=13$ positive relevant unknowns read:
\begin{description}
\item $\Prob(1)$, $\Prob( 2)$,  $\Prob( 3)$,  $\Prob( 4)$,  
$\Prob( 5)$, $\Prob( 6)$, $\Prob( 7)$, $\Prob( 8)$, 
\item $\Prob(  1; 3)$, $\Prob(  4; 7)$, $\Prob(  2; 7)$, $\Prob(  2; 4)$, 
\item $ \Prob(  2; 4; 7)$.  
\end{description}

\paragraph{Relevant universal equations}

We are now going to account for the laws of logic. According to consistency theorems by Richard Cox~\cite{cox} these laws are expressed in the quantitative rules of probability theory~\cite{jaynes}. In the present context, they give rise to a number of consistency constraints that we have called \emph{universal equations}. They are conveniently derived from the list of all relevant positive unknowns.

- For each  relevant positive unknown of one literal $\Prob(k)$ we have one normalization equation:
\begin{align}
\label{simplet}
1=\Prob(k)+\Prob(-k)
\end{align} 
that is a total of $4n$ equations.

- For each  relevant positive unknown of two literals $\Prob(k_1;k_2)$, we have $4$ consistency equations
\begin{align}
\label{doublet}
\begin{aligned} 
\Prob(\pm k_1)=\Prob(\pm k_1;k_2)+\Prob(\pm k_1; -k_2)\\
\Prob(\pm k_2)=\Prob(\pm k_2;k_1)+\Prob(\pm k_2; -k_1)
\end{aligned}
\end{align}
that is a total of $4\times(3n-2)=12n-8$ equations.

- For each  relevant positive unknown of three literals $\Prob(k_1;k_2;k_3)$, we have $3\times 4=12 $ consistency equations, namely
\begin{align}
\label{triplet}
\begin{aligned} 
\Prob(\pm k_1;\pm k_2)= \Prob(\pm k_1;\pm k_2 ; k_3) +  \Prob(\pm k_1;\pm k_2 ; -k_3) \\
\Prob(\pm k_2;\pm k_3)= \Prob(\pm k_2;\pm k_3; k_1) + \Prob(\pm k_2;\pm k_3 ; -k_1) \\
\Prob(\pm k_3;\pm k_1)= \Prob(\pm k_3;\pm k_1 ; k_2) + \Prob(\pm k_3;\pm k_1 ; -k_2) 
\end{aligned}
\end{align} 
that is a total of $12\times(n-1)$  equations.
 
Collecting these results, the numbers of relevant universal equations are the following:
\begin{center}
\begin{tabular}{||c|c|c||}
\hline
\hline
literals& positive unknowns     & universal equations  \\
\hline
1& $4n$                    & $4n$ \\
2& $3n-2$                  & $12n-8$\\
3& $n-1$                  & $12n-12$ \\
Total& $8n-3$                 & $28n-20$ \\
\hline
\hline
\end{tabular}
\end{center}

\emph{Example:} Let again $n=2$. 
We have 

- $4n=8$  normalization equations for $8$ positive literals, namely
\begin{align}
\begin{aligned}
\label{n2univ1}
1=\Prob(1)+\Prob(-1)\ ; 
1=\Prob(2)+\Prob(-2)\ ;
1=\Prob(3)+\Prob(-3)\ ;
1=\Prob(4)+\Prob(-4)\\
1=\Prob(5)+\Prob(-5)\ ;
1=\Prob(6)+\Prob(-6)\ ;
1=\Prob(7)+\Prob(-7)\ ;
1=\Prob(8)+\Prob(-8).
\end{aligned}
\end{align} 

- $4\times (3n-2) = 16$ universal equations corresponding to  $3n-2=4$ positive unknowns with 2 literals. 
\begin{align}
\label{n2univ2}
\begin{aligned}
\Prob(\pm 1)=\Prob(\pm 1;3)+\Prob(\pm 1; -3) ;\quad
\Prob(\pm 3)=\Prob(\pm 3;1)+\Prob(\pm 3; -1) \\
\Prob(\pm 2)=\Prob(\pm 2;7)+\Prob(\pm 2; -7) ;\quad
\Prob(\pm 2)=\Prob(\pm 2;4)+\Prob(\pm 2; -4) \\
\Prob(\pm 4)=\Prob(\pm 4;7)+\Prob(\pm 4; -7) ;\quad
\Prob(\pm 4)=\Prob(\pm 4;2)+\Prob(\pm 4; -2)\\
\Prob(\pm 7)=\Prob(\pm 7;4)+\Prob(\pm 7; -4) ;\quad
\Prob(\pm 7)=\Prob(\pm 7;2)+\Prob(\pm 7; -2) \\
\end{aligned}
\end{align} 

- $12n-12 = 12$ universal equations for the single 3-literal positive unknown,
\begin{align}
\label{n2univ3}
\begin{aligned}
\Prob(\pm 2;\pm 4)= \Prob(\pm 2;\pm 4 ; 7) +  \Prob(\pm 2;\pm 4 ; -7) \\
\Prob(\pm 4;\pm 7)= \Prob(\pm 4;\pm 7 ; 2) + \Prob(\pm 4;\pm 7 ; -2) \\
\Prob(\pm 2;\pm 7)= \Prob(\pm 2;\pm 7 ; 4) + \Prob(\pm 2;\pm 7 ; -4) \\
\end{aligned}
\end{align} 
that is a total of 
$28n-20=36$ relevant universal equations. The total number of equations is then $(36$ universal equations$) + (8$ specific equations$)= 44$ equations.

\paragraph{Linear programming implementation} 
At this point, we have completed the conversion of all Boolean formulae into linear equations operating in a real vector space of the \emph{unknown}-vectors with $N=28n-16$ dimensions. Thanks to the set of universal equations, these unknowns can be consistently interpreted as partial probabilities in a Bayesian probability space.  Therefore they are non negative. This defines a linear programming problem~\cite{algo, murty} which can be solved in polynomial time in $N=$ O$(n)$~\cite{khachiyan}.
If the number of data bits is $n_b$, we have a total of $n_b+30n-20$ linear equations, e.g.,  $32n-20$ for the conventional addition. 

Due to the product rule  in the probability space we have:
\begin{align}
\label{probaRule}
\begin{aligned}
&\Prob(k_1)  =   0 \Rightarrow \Prob(k_1;k_2) =  0 \Rightarrow \Prob(k_1;k_2;k_3) =  0. \\
&\Prob(k_1) =1\Rightarrow \Prob(k_1;k_2) =  \Prob(k_2)  \\
&\Prob(k_1) =1\Rightarrow \Prob(k_1;k_2;k_3) =  \Prob(k_2;k_3)  \\
\end{aligned}
\end{align}
It is convenient to take advantage of these relations to simplify the linear system by accounting beforehand for the data equations. As a result, a number of equations of the rough system are cancelled and a number of unknowns become irrelevant. 

Generally, only \emph{deterministic solutions} are of interest.
Note that deterministic solutions are also \emph{separable}~\cite{mf3}, i.e., the probability of any requirement is the product of the probabilities of its literals, e.g.,
\begin{equation}
\label{separable}
\Prob(k_1;k_2;k_3) =  \Prob(k_1)\times\Prob(k_2)\times\Prob(k_3).
\end{equation}

\emph{When the LP problem is feasible}, 
we obtain a value for each relevant partial probability. Deterministic solutions are in principle computed by LP-optimization in a time O$(n)$.  However, in this simple case of addition, it turns out that a feasible problem always accept a deterministic solution and even  optimization is unnecessary when accounting for Eq.(\ref{probaRule}). The conclusion will be different for multiplication.

\emph{When the LP problem is not feasible}, the problem has no solution, e.g., $S-U$ when $S<U$. 

\paragraph{}

\emph{Example 1:} Let $n=1$.  Consider the addition $S=U+V$ given that $U=0$ and $V=1$. We have $12$ unknowns and $32n-20=12$ equations in the rough system: 
\begin{align}
\label{add1bit}
\begin{aligned}
&\Prob(1) =  0\ ;\quad
\Prob(-2) =  0\ ; \\
&\Prob(3) = \Prob(2;-1) + \Prob(-2;1) =  0\ ;\quad
\Prob(4) = \Prob(2;1)\ ;  \\
&\Prob(-1) + \Prob(1) =  1\ ;\quad
\Prob(-2) + \Prob(2) =  1\ ;\\
&\Prob(-3) + \Prob(3) =  1\ ;\quad
\Prob(-4) + \Prob(4) =  1 \ ; \\
&\Prob(2) = \Prob(2;-1) + \Prob(2;1) ;  \quad
\Prob(-2) = \Prob(-2;-1) + \Prob(-2;1)\ ;\\
&\Prob(1) = \Prob(-2;1) + \Prob(2;1)\ ;\quad
\Prob(-1) = \Prob(-2;-1) + \Prob(2;-1).
\end{aligned} 
\end{align}
The rank of the system is $11$. When accounting for Eq. (\ref{probaRule}), two equations become identically zero and thus the four unknowns with more than one literal become irrelevant. We have then,
\begin{align*}
\begin{aligned}
&\Prob(1) =  0 \ ; \
\Prob(-2) =  0\ ; \\
&\Prob(3) = \Prob(-1) \ ; \
\Prob(4) =  0 \ ; \\
&\Prob(-1) =  1 \ ; \
\Prob(2) =  1\ ; \
\Prob(-3) + \Prob(3) =  1\ ; \
\Prob(-4) + \Prob(4) =  1\ ; \
\Prob(-1) = \Prob(2).
\end{aligned} 
\end{align*}%
The first row is just the probability formulation of the data: $U=0\ ; V=1$. The second row described the two structural equations. The last row displays the universal equations. The resolution is straightforward. 
The rank of the linear system is now $8=8n$ and the LP problem is feasible with the conventional deterministic solution $S=1$.

\paragraph{}

\emph{Example 2:} Let $n=2$. Consider the addition $S = U+V$ given $U=2$ and $V=3$. We have $28n-16 = 40$ unknowns with $44$ equations, namely Eqs. (\ref{n2specificplus}, \ref{n2univ1}, \ref{n2univ2}, \ref{n2univ3}). The rank of the rough linear system is $35$. The LP systeme is feasible. Using Eq.(\ref{probaRule}), the rank is at once lowered to $16=8n$ and all unknowns with more than one literal become irrelevant. After Gauss elimination the matrix is diagonal, so no LP algorithm is needed. This result seems general: The system is trivial when accounting for Eq.(\ref{probaRule}). 

\paragraph{}

\emph{Example 3:} Let $n=1$.  Consider the subtraction $V=S-U$ given that $S=0$ and $U=1$. The linear system Eq.(\ref{add1bit}) is still valid, except the first row which is now $\Prob(-1) =  0 ; \Prob(3) =  0 ;\Prob(4) =  0$. 
We have still $12$ unknowns but $32n-19=13$ equations. The rank of the linear system is $12$ and the single solution is
\begin{align*}
    &\Prob(-1)  =   0 \ ;\
  &&\Prob(-2)  =   2 \ ;\
&\Prob(-3)  =   1 \ ;\
   &&\Prob(-4)  =   1 \ ;\
 &\Prob(-2;1)  =   1 \ ;\
   &&\Prob(-2;-1)  =   1 \ ; \\
   &\Prob(1)  ~~ =   1 \ ;\
   &&\Prob(2)  =  -1 \ ;\
   &\Prob(3)  =   0 \ ;\
    &&\Prob(4)  =   0 \ ;\
   &\Prob(2;1)  ~~~=   0 \ ;\
 &&\Prob(2;-1)  =  -1. \\
\end{align*}
The LP systeme is not feasible because e.g. $ \Prob(2)  <0$. When accounting for Eq.(\ref{probaRule}) the linear system is at once impossible. However, in general when $S<U$ the linear system is possible but the LP problem is not feasible.

\paragraph{}

\emph{Addition of several integers}: The addition of {several integers} can be performed step by step. We start with an integer $U_0$. We add a first term $U_1$ to obtain a first sum $S_1$ and internal variables for the carry bits. Next, we add the second term $U_2$ to $S_1$, etc. We will use this process in the next section for the Bayesian multiplication.

\paragraph{}

\emph{Remark on the Bayesian addition.}
If we just have to add or subtract two integers, the Bayesian environment is certainly not a paragon of simplicity. For instance, we have seen that the sum of two integers of two bits is solved by a LP problem with $40$ unknowns and $44$ equations! The same conclusion can be drawn for any operation known to be of complexity \textsf{P}. Nevertheless, for exotic problems of the addition environment, if we have an input of $n_b$ data bits and no algorithm except a \emph{force brute} exploration of the  $2^{4n-n_b}$ potential assignments, the method will be useful for large $n$. For what we are concerned, we will use these results in a more complex environment, namely, multiplication. 


\section {Multiplication}
We now aim to encode the product of two integers $A$ and $B$ in the framework of Bayesian arithmetic. Most of the ingredients are directly derived from the previous section and thus we will just have to replicate the results with a minimum of details. Let $n$ and $m$ respectively be the number of bits of the two integers. 
When setting down a conventional binary multiplication we have simply to add $m$ terms, $U_0$, $U_1$,\dots, $U_{m-1}$. The addition will be processed in $m-1$ steps. In this context, the integers are `shifted', that is 
$U_0$ is composed of $n$ bits and the following terms  $U_t$ are also composed of $n$ significant bits followed by $t$ zeros. 

\subsection{Addition of shifted integers}

Let $t\in\llbracket 0,m-1\rrbracket$. We have 
$$U_t=  \sum_{i=0}^{n-1} u_{t, t+i} 2^{t+i} $$

\emph{Example} Let us write the binary additions of shifted integers for $n=2$ and $m=3$. Let $r_{t,t+i}$ be the carry bits.
\begin{center}
\begin{tabular}{L|CCCCCC}
U_0 &&                 &          &              & \bf{u_{01}} &   \bf{u_{00}} \\
U_1 &&                &           &  \bf{u_{12}}   & \bf{u_{11}} &       .    \\
R_1 &&               & \bf{r_{13}}  &  \bf{r_{12}}   &         .  &        .    \\
\hline
S_1=U_0+U_1 &&               & {r_{13}}  &  \bf{s_{12}}   & \bf{s_{11}} & u_{00} \\
U_2 &&               & \bf{u_{23}}  &  \bf{u_{22}}   & .          &  . \\
R_2 &&      \bf{r_{24}}         & \bf{r_{23}}  &  .   &         .  &      .      \\
\hline
S_2 = S_1+U_2& &  {r_{24}}   & \bf{s_{23}}  &  \bf{s_{22}}   &  s_{11}  &   u_{00}   \\
\end{tabular}
\end{center}

We note that we have $n$ significant bits for each integer $U_0$,  $U_t$, $S_t$ and a `carry integer' $R_t$ for $t\in\llbracket 1,m-1\rrbracket$ (displayed in boldface). 
Let $\mathsf{U}_{0,i}$, $\mathsf{U}_{t,t+i}$, $\mathsf{S}_{t,t+i}$ and $\mathsf{R}_{t,t+i+1}$ be the random variables describing these integers for $i\in\llbracket 0,n-1\rrbracket$.
Again, it is suitable to bring together these $3nm-2n$ variables in a single list $\mathsf{X}_k$ of \emph{global variables}, labelled from $k=1$ to $3nm-2n$. We will adopt the labelling convention of table~\ref{varglobmult}. 
\begin{table}
\footnotesize{
\begin{center}
\begin{tabular}{||C|C|C|C||}
\hline
\hline
\mathsf{U}_{0,i} &\mathsf{U}_{t,i}  &  \mathsf{S}_{t,i}  & \mathsf{R}_{t,i} \\
\hline
i\in\llbracket 0, n-1\rrbracket &  i\in\llbracket t, t+n-1\rrbracket &  i\in\llbracket t, t+n-1\rrbracket & i\in\llbracket t+1, t+n\rrbracket \\
\hline
 k=i+1&  k=n(3t-2)+i-t+1& k=n(3t-1)+i-t+1 & k=3nt+i-t\\
\hline
k\in\llbracket 1, n\rrbracket &  k\in\llbracket (3t-2)n+1, (3t-1)n\rrbracket &  k\in\llbracket (3t-1)n+1, 3tn\rrbracket & k\in\llbracket 3tn+1, n(3t+1)\rrbracket \\
\hline
 \mathsf{X}_{1} \mathrm{~to~} \mathsf{X}_{n}   &\mathsf{X}_{(3t-2)n+1} \mathrm{~to~} \mathsf{X}_{(3t-1)n}   &  \mathsf{X}_{(3t-1)n+1}\mathrm{~to~} \mathsf{X}_{3tn}  & \mathsf{X}_{3tn+1} \mathrm{~to~} \mathsf{X}_{n(3t+1)} \\
\hline
\hline
\end{tabular}
\caption{\footnotesize
\label{varglobmult}
Labelling convention of the global variables $\mathsf{X}_k$ corresponding to the addition of $m$ shifted integers of $n$ bits. We have $t\in\llbracket 1, m-1\rrbracket$ and a total of  $3nm-2n$ variables.}
\end{center}
}
\end{table}

\emph{Example} When $n=2$ and $m=3$ the shifted addition reads, 

\begin{center}
\begin{tabular}{L|CCCCCC}
U_0 &&                 &          &              &  \mathsf{X}_{\bf{2}}&    \mathsf{X}_{\bf{1}}\\
U_1 &&                &           &   \mathsf{X}_{\bf{4}}&  \mathsf{X}_{\bf{3}}&       .    \\
R_1 &&               & \mathsf{X}_{\bf{8}}&  \mathsf{X}_{\bf{7}}&         .  &        .    \\
\hline
S_1=U_0+U_1 &&               & \mathsf{X}_{\bf{8}}  &   \mathsf{X}_{\bf{6}}&  \mathsf{X}_{\bf{5}}& \mathsf{X}_{\bf{1}}\\
U_2 &&               &  \mathsf{X}_{\bf{10}}&   \mathsf{X}_{\bf{9}}& .          &  . \\
R_2 &&       \mathsf{X}_{\bf{14}}&  \mathsf{X}_{\bf{13}}&  .   &         .  &      .      \\
\hline
S_2 = S_1+U_2& &  \mathsf{X}_{\bf{14}}   &  \mathsf{X}_{\bf{12}}&   \mathsf{X}_{\bf{11}}&  \mathsf{X}_{\bf{5}}  &   \mathsf{X}_{\bf{1}}   \\
\end{tabular}
\end{center}

\paragraph{Structural equations}
Since we proceed with the computation step by step,  we just have to bring together the structural equations of each step. We extract from the full operation the relevant computation section and apply the previous result concerning the addition of two integers.


\emph{First step:} For $t=1$, the relevant computation section is the following,

\begin{center}
\begin{tabular}{L|CCCCCCCC}
U_0& &&\bf{u_{0,n-1}} & \dots& \bf{u_{0,i}}      &\dots&  \bf{u_{0,2}}   & \bf{u_{0,1}}        \\
U_1 &&\bf{u_{1,n}}&\bf{u_{1,n-1}}& \dots&  \bf{u_{1,i}} &\dots&    \bf{u_{1,2}}             & \bf{u_{1,1}}     \\
R_1 &\bf{r_{1,n+1}}& \bf{r_{1,n}}&\bf{r_{1,n-1}}&\dots&   \bf{r_{1,i}}     &\dots &     \bf{r_{1,2}}         & .   \\
\hline
S_1 &{r_{1,n+1}} &\bf{s_{1,n}} &  \bf{s_{1,n-1}}   & \dots &   \bf{s_{1,i}}  &\dots&  \bf{s_{1,2}}  & \bf{s_{1,1}}   \\
\end{tabular}
\end{center}

\emph{Next steps: } For $t\in\llbracket 2,m-1 \rrbracket$, the computation section reads,

\begin{center}
\begin{tabular}{L|CCCCCCC}
S_{t-1} &&{r_{t-1,t+n-1}}& \dots& \bf{s_{t-1,t+i}}      &\dots&  \bf{s_{t-1,t+1}}   & \bf{s_{t-1,t}}        \\
U_t &&\bf{u_{t,t+n-1}}& \dots&  \bf{u_{t,t+i}}        &\dots&    \bf{u_{t,t+1}}             & \bf{u_{t,t}}     \\
R_t & \bf{r_{t,t+n}}&\bf{r_{t,t+n-1}}&\dots&   \bf{r_{t,t+i}}     &\dots &     \bf{r_{t,t+1}}         & .   \\
\hline
S_t & {r_{t,t+n}}   &  \bf{s_{t,t+n-1}}   & \dots &   \bf{s_{t,t+i}}  &\dots&  \bf{s_{t,t+1}}  & \bf{s_{t,t}}   \\
\end{tabular}
\end{center}

Collecting the results of each step, be obtain the set of structural equations expressed in terms of global variables given in  table \ref{equamult}. We have $2n$ equations per section and $m-1$ sections, i.e., a total of $2n(m-1)$ structural equations.

\begin{table}
\footnotesize{
\fbox{
\begin{minipage}{\textwidth}
\begin{align*}
\Prob(2n+1)&=\Prob(-2 ;n+1) + \Prob(2 ;-n-1)\\
\Prob(3n+1)&= \Prob(2 ;n+1)\\
\Prob({2n+i})&= \Prob({i+1}; {-i-n};{-i-3n+1}) +\Prob({-i-1}; {i+n};{-i-3n+1}) 
+ \Prob({-i-1}; {-i-n};{i+3n-1}) + \Prob({i+1}; {i+n};{i+3n-1}) \\
 \Prob({3n+i})&=  \Prob({i+1}; {i+n};{-i-3n+1}) 
+\Prob({-i-1}; {i+n};{i+3n-1}) 
+ \Prob({i+1}; {-i-n};{i+3n-1}) + \Prob({i+1}; {i+n};{i+3n-1}) \\
\Prob({3n})&=  \Prob({2n}; {-4n+1})+ \Prob({-2n}; {4n-1}) \\
 \Prob({4n})&=  \Prob({2n}; {4n-1}) \\
\Prob(3nt-n+1)&=\Prob(3nt-4n+2 ; -3nt+2n-1) + \Prob(-3nt+4n-2;3nt-2n+1)\\
\Prob(3nt+1)&= \Prob(3nt-4n+2 ;3nt-2n+1).\\
\Prob({3nt-n+i})&= \Prob(-3nt+4n-i-1; -3nt+2n-i ;3nt+i-1) 
+\Prob({-3nt+4n-i-1}; {3nt-2n+i};{-3nt-i+1}) \\
&\quad\quad\quad\quad+ \Prob({3nt-4n+i+1}; {-3nt+2n-i};{-3nt-i+1}) 
+ \Prob({3nt-4n+i+1}; {3nt-2n+i};{3nt+i-1}) \\
 \Prob({3nt+i})&=  \Prob(3nt-4n+i+1; 3nt-2n+i ;-3nt-i+1) 
+\Prob({3nt-4n+i+1}; {-3nt+2n-i};{3nt+i-1}) \\
&\quad\quad\quad\quad+ \Prob({-3nt+4n-i-1}; {3nt-2n+i};{3nt+i-1}) 
+ \Prob({3nt-4n+i+1}; {3nt-2n+i};{3nt+i-1}) \\
\Prob({3nt})&= \Prob(-3nt+2n; -3nt+n ;3nt+n-1)
+\Prob({-3nt+2n}; {3nt-n};{-3nt-n+1}) \\
&\quad\quad\quad\quad + \Prob({3nt-2n}; {-3nt+n};{-3nt-n+1}) 
+ \Prob({3nt-2n}; {3nt-n};{3nt+n-1}) \\
 \Prob({3nt+n})&=  \Prob(3nt-2n; 3nt-n ;-3nt-n+1) 
+\Prob({3nt-2n}; {-3nt+n};{3nt+n-1}) \\
&\quad\quad\quad\quad + \Prob({-3nt+2n}; {3nt-n};{3nt+n-1}) 
+ \Prob({3nt-2n}; {3nt-n};{3nt+n-1}) \\
\end{align*}
\end{minipage}
}
\caption{\footnotesize
\label{equamult}
Structural equations of the addition of $m$ shifted integers of $n$ bits.
($i \in\llbracket 2, n-1\rrbracket$ ; $t\in\llbracket 2,m-1\rrbracket$). We have a total of $2n(m-1)$ equations.
}
}
\end{table}

\emph{Example:} For $n=2$ and $m=3$,  we have $2n(m-1)= 8$ structural equations:
\begin{align*}
\Prob( 5 )&=\Prob(-2 ; 3 ) + \Prob( 2 ;-3 )\\
\Prob( 7 )&= \Prob( 2  ; 3 )\\
\Prob( 6 )&=  \Prob( 4 ; -7 )+ \Prob(-4 ;  7 ) \\
\Prob( 8 )&=  \Prob( 4 ;  7 ) \\
\Prob( 11 )&=\Prob( 6 ;-9 ) + \Prob(-6 ; 9 )\\
\Prob( 13 ) &= \Prob( 6 ; 9 )\\
\Prob( 12 )&= \Prob(-8 ;-10 ; 13 ) +\Prob(-8 ;  10 ;-13 ) + \Prob( 8 ; -10 ;-13 ) + \Prob( 8 ;  10 ; 13 ) \\
\Prob( 14 )&=  \Prob( 8 ; 10 ;-13 )+\Prob( 8 ; -10 ; 13 ) + \Prob(-8 ;  10 ; 13 ) + \Prob( 8 ;  10 ; 13 ) \\
\end{align*}
%

\paragraph{Relevant unknowns}
In order to list the relevant unknowns, we inspect the structural equations. We have $(mn-m-n)$ 3-literal positive unknowns and then $3\times(mn-m-n)$ 2-literal positive variants. We have $m$ direct 2-literal positive unknowns and finally $(3mn-2n)$ 1-literal positive unknowns. Gathering together  these results we have a total $7mn-3m-6n$ positive unknowns and a total $26mn-16m-24n$ unknowns (Table \ref{posiunknowns}).

\begin{center}
\begin{tabular}{||c|c|c||}
\hline
\hline
literals& positive unknowns     & unknowns \\
\hline
1  & $3mn    -2n$          & $6mn        -4n  $  \\
2  & $3mn-2m-3n$        & $12mn -8m-12n $  \\
3  & $mn  -m  -n$          & $8mn  -8m -8n  $  \\
Total   & $7mn-3m-6n$         & $26mn-16m-24n$ \\
\hline
\hline
\end{tabular}
\end{center}

\begin{table}
\footnotesize{
\fbox{
\begin{minipage}{\textwidth}
\begin{align*}
&\Prob(k)\\
&\Prob(2 ;n+1) ; \Prob(2n; 4n-1)\ ;\ \Prob(3nt-4n+2 ;3nt-2n+1)\\
&\Prob( {i+n};{i+3n-1}) \ ;\  \Prob({i+1}; {i+3n-1}) \ ;\  \Prob({i+1}; {i+n})\\
&\Prob( {3nt-2n+i};{3nt+i-1}) \ ;\  \Prob({3nt-4n+i+1}; {3nt+i-1}) \ ;\  \Prob({3nt-4n+i+1}; {3nt-2n+i})\\
&\Prob( {3nt-n};{3nt+n-1}) \ ;\  \Prob({3nt-2n}; {3nt+n-1}) \ ;\ \Prob({3nt-2n}; {3nt-n})\\
&\Prob({i+1}; {i+n};{i+3n-1}) \ ;\ \Prob({3nt-2n}; {3nt-n};{3nt+n-1})\\
&\Prob({3nt-4n+i+1}; {3nt-2n+i};{3nt+i-1})\\
\end{align*}
\end{minipage}
}
\caption{\footnotesize
\label{posiunknowns}
Positive unknowns of the addition of $m$ shifted integers of $n$ bits.
($i \in\llbracket 2, n-1\rrbracket$ ;
$t\in\llbracket 2,m-1\rrbracket$ ;
$k\in\llbracket 1,n(3m-2)\rrbracket$). We have a total of $7mn-3m-6n$ positive unknowns.
}
}
\end{table}

\emph{Example:} Let $n=2$ and $m=3$, we have $7mn-3m-6n = 21$ relevant positive unknowns,
\begin{description}
\item $\Prob( k ), k = 1$ to $14$
\item $\Prob( 2 ; 3 )$, $\Prob( 7 ; 4 )$, $\Prob( 6 ; 9 )$, 
$\Prob( 10 ; 13 )$, $\Prob( 8 ;  13 )$, $\Prob( 8 ; 10  )$
\item $\Prob( 8 ; 10 ; 13 )$
\end{description}

\paragraph{Relevant universal equations} 
The relevant universal equations are derived from the relevant positive unknowns by Eq.(\ref{simplet}, \ref{doublet}, \ref{triplet}).
 \begin{center}
\begin{tabular}{||c|c|c||}
\hline
\hline
literals& positive unknowns   & universal equations  \\
\hline
1              & $3mn    -2n$          &     $3mn           -2n$          \\
2             & $3mn-2m-3n$        &    $12mn     -8m -12n$                \\
3             & $mn  -m  -n$           &    $12mn   -12m -12n$      \\
Total       & $7mn-3m-6n$          & $27mn-20m- 24n$ \\
\hline
\hline
\end{tabular}
\end{center}
For instance, for $n=2$ et $m=3$, we have $27mn-20m- 24n = 54$ relevant universal equations.

We have completed the analysis of the addition of $m$ shifted integers of $n$ bits. We have identified $26mn-16m-24n$ relevant unknowns,  $2n(m-1)$ structural equations and $27mn-20m- 24n$ universal equations, but not defined any data equation.
We are now ready to apply these results to the very multiplication.


\subsection{Multiplication}
Let  $A$ and $B$  be two integers with binary expansions,
$$ A = \sum_{i=0}^{n-1} 2^i a_i \quad ; \quad B = \sum_{t=0}^{m-1} 2^t b_t, $$
where $n$ and $m$ are the number of bits of $A$ and $B$ respectively. Let $C$ be a third integer and $\Pi$ the logical proposition
$$ \Pi : \quad A\times B = C. $$
We suppose throughout that $\Pi$ is satisfied. The binary expansion of $C$ reads
$$C =\sum_{i=0}^{n-1}\sum_{t=0}^{m-1}2^{i+t} a_i b_t= \sum_{j=0}^{n+m-1} 2^j c_j $$
For $t\in\llbracket 0, m-1\rrbracket$, define $U_t$,  
\begin{align*}
U_t &= \sum_{i=0}^{n-1} 2^{i+t} a_i b_t =  \sum_{i=t}^{t+n-1} u_{t,i} 2^i 
\quad\mathrm{~so~that~}\quad
C= \sum_{t=0}^{m-1} U_t
\end{align*}
Clearly, the integers $U_t$ form a set of $m$ shifted integers as described in the previous section. Thus, we will take back the random variables $\mathsf{U}_{t,i}$, $\mathsf{S}_{t,i}$, $\mathsf{R}_{t,i}$. We will also define new random variables $\mathsf{A}_{i}$ and $\mathsf{B}_{t}$ corresponding to the binary expansion of $A$ and $B$. The binary variables of $C$ are already ranked because $C= S_{m-1}$. Eventually, we construct a probability space with all these variables, namely $\mathsf{U}_{t,i}$, $\mathsf{S}_{t,i}$, $\mathsf{R}_{t,i}$, $\mathsf{A}_{i}$ and $\mathsf{B}_{t}$, and define a Bayesian probability distribution $\Prob$ given $\Pi$. 

It is necessary to extend the list of global variables $\mathsf{X}_k$, (Table \ref{varglobmult}), in order to account for $A$ and $B$. We will adopt the convention of Table \ref{varfactors}.  The global variables corresponding to the bits of $C= S_{m-1}$  are recalled in Table \ref{varproduit}.

\begin{table}
{\footnotesize
\begin{center}
\begin{tabular}{||C|C||}
\hline
\hline
\mathsf{A}_0 \mathrm{~to~} \mathsf{A}_{n-1} &  \mathsf{B}_0 \mathrm{~to~} \mathsf{B}_{m-1}  \\
\hline
i\in\llbracket 0,n-1\rrbracket & t\in\llbracket 0,m-1\rrbracket \\
\hline
 k=i+3nm-2n+1& k=t+3nm-n+1 \\
\hline
k\in\llbracket 3nm-2n+1,3nm-n\rrbracket & k\in\llbracket 3nm-n+1,3nm-n+m\rrbracket \\
\hline
 \mathsf{X}_{3nm-2n+1}  \mathrm{~to~} \mathsf{X}_{3nm-n} &  \mathsf{X}_{3nm-n+1} \mathrm{~to~} \mathsf{X}_{3nm-n+m}   \\
\hline
\hline
\end{tabular}
\end{center}
\caption{\footnotesize
\label{varfactors}
Labelling convention of the global variables $\mathsf{X}_k$ corresponding to the variables $\mathsf{A}_i$, $\mathsf{B}_t$. This defines a number of $mn$ variables.}
}
\end{table}
\begin{table}
~\\
\footnotesize{
\begin{center}
\begin{tabular}{||C|C|C|C||}
\hline
\hline
\mathsf{C}_0  &  \mathsf{C}_1 \mathrm{~to~} \mathsf{C}_{m-1} &  \mathsf{C}_{m} \mathrm{~to~} \mathsf{C}_{m+n-2} & \mathsf{C}_{m+n-1}\\
\hline
j=0 & j\in\llbracket 1,m-1\rrbracket & j\in\llbracket m,m+n-2\rrbracket & j=m+n-1 \\
\hline
k=1& k=3nj-n+1& k=3nm-4n-m+2+j& k=3nm-2n\\
\hline
 \mathsf{X}_{1}   &  \mathsf{X}_{2n+1} \mathrm{~to~} \mathsf{X}_{3nm-4n+1}  &  \mathsf{X}_{3nm-4n+2} \mathrm{~to~} \mathsf{X}_{3nm-3n}  &  \mathsf{X}_{3nm-2n} \\
\hline
\hline
\end{tabular}
\end{center}
\caption{\footnotesize
\label{varproduit}
Labelling convention of the global variables $\mathsf{X}_k$ corresponding to the variables $\mathsf{C}_j$, already defined in Table \ref{varglobmult} to describe the
 variables $\mathsf{U}_{0,0}$, $\mathsf{S}_{t,t}$, $\mathsf{S}_{m-1,j}$ and $\mathsf{R}_{m-1,m+n-1}$.}
}
\end{table}

\emph{Example} Let $n=2$ and $m=3$. We have
\begin{description}
\item $\mathsf{A}_{0} = \mathsf{X}_{15}$ ;
$\mathsf{A}_{1} = \mathsf{X}_{16}$.
\item 
$\mathsf{B}_{0} = \mathsf{X}_{17}$ ;
$\mathsf{B}_{1} = \mathsf{X}_{18}$ ;
$\mathsf{B}_{2} = \mathsf{X}_{19}$ ;
\item
$\mathsf{C}_{0} = \mathsf{X}_{1}$ ;
$\mathsf{C}_{1} = \mathsf{X}_{5}$ ;
$\mathsf{C}_{2} = \mathsf{X}_{11}$ ;
$\mathsf{C}_{3} = \mathsf{X}_{12}$ ;
$\mathsf{C}_{4} = \mathsf{X}_{14}$ ;
\end{description}

\paragraph{Structural equations}
In addition to the structural equations of Table~\ref{equamult}, let us consider 
the truth table of the variables $\mathsf{U}_{t,t+i}$ versus $\mathsf{A}_{i}$ and $\mathsf{B}_{t}$.
\begin{center}
\begin{tabular}{||C|C||C||}
\hline
\hline
\mathsf{B}_t&\mathsf{A}_i&\mathsf{U}_{t,t+i}\\
\hline
0&0&0\\
0&1&0\\
1&0&0\\
1&1&1\\
\hline
\hline
\end{tabular}
\end{center}
The codification of this truth table into linear equations is straightforward.
$$\Prob(\mathsf{U}_{t,t+i}=1|\Pi)= \Prob(\mathsf{A}_{i}=1 ; \mathsf{B}_{t}=1|\Pi)$$
Thus, we obtain a set of $mn$ new structural equations displayed with the global variable convention in Table~\ref{equaproduit}, to be  added to the $2mn-2n$ equations of table \ref{equamult} and leading to a total of $3mn-2n$ structural equations. 

\begin{table}
\footnotesize{
\fbox{
\begin{minipage}{\textwidth}
\begin{align*}
\Prob(i) &= \Prob(i+3nm-2n+1 ; 3nm-n+m+3)\\
\Prob(n(3t-2)+i-t+1)& = \Prob(i+3nm-2n+1 ; t+3nm-n+m+3)\\
\end{align*}
\end{minipage}
}
\caption{\footnotesize
\label{equaproduit}
Complementary structural equations of the multiplication ($ i\in\llbracket 0,n-1\rrbracket$,  $ t\in\llbracket 1,m-1\rrbracket$) in addition to Table~\ref{equamult}. The number of new equations is $mn$ and the total number of structural equations is $ 3nm - 2n$.}
}
\end{table}

\paragraph{Relevant unknowns}

Again, we take back the relevant unknowns of the shifted addition (Table~\ref{posiunknowns}). We have to add $n+m$ new positive relevant unknowns of one literal, namely $\Prob(\mathsf{A}_i|\Pi)$ and $\Prob(\mathsf{B}_t|\Pi) $ and $nm$  new positive relevant unknowns of two literals, namely $\Prob(\mathsf{A}_i;\mathsf{B}_t|\Pi)$. They are listed in Table~\ref{posifact}.

\begin{table}
\footnotesize{
\fbox{
\begin{minipage}{\textwidth}
\begin{align*}
&\Prob(k)\\
&\Prob(3nm-2n+1 +i;3nm-n+1+t)
\end{align*}
\end{minipage}
}
\caption{\footnotesize
\label{posifact}
Complementary positive unknowns of the multiplication in addition to Table~\ref{posiunknowns}.
($i \in\llbracket 0, n-1\rrbracket$ ;
$t\in\llbracket 0,m-1\rrbracket$ ;
$k\in\llbracket 3nm-2n+1,3nm-n+m\rrbracket$). 
We have $mn+m+n$ complementary positive unknowns and a total of $8mn-2m-5n$ positive unknowns.
}
}
\end{table}

\emph{Example}: Let $n=2$ and $m=3$. The 11 new relevant positive unknowns are,
\begin{description}
\item  \Prob(15) ; \Prob(16) ; \Prob(17); \Prob(18) ; \Prob(19)
\item  \Prob(19;15) ; \Prob(19;16) ; \Prob(18;15) ; \Prob(18;16) ; \Prob(17;15) ; \Prob(17;16) 
\end{description}

The unknowns are derived from the positive unknowns from enumeration of the variants.
\begin{center}
\begin{tabular}{||c|c|c|c||}
\hline
\hline
literal& positive unknowns      & unknowns  & universal equations\\
\hline
1               & $3mn+m  -n$          & $6mn +2m       -2n  $ & $3mn+m  -n$\\
2             & $4mn-2m-3n$          & $16mn -8m-12n $  & $   16mn -8m-12n $  \\
3             & $mn  -m  -n$           & $8mn  -8m -8n  $  & $     12mn  -12m  -12n$           \\
Total       & $8mn-2m-5n$         & $30mn-14m-22n$ &   $31mn -19m-25$                                     \\
\hline
\hline
\end{tabular}
\end{center}

\paragraph{Relevant universal equations} We derive the relevant universal equations from the list of positive unknowns. We have a total of  $ 1\times (3nm-n+m) + 4 \times (4mn-2m-3n) + 12\times (mn-m-n) = 31mn-19m-25n$ relevant universal equations.

In summary, for $m, n>1$ we have

\begin{center}
\begin{tabular}{lL}
unknowns: & 30mn-14m-22n\\
structural equations: &3mn-2n\\
universal equations: &31mn-19m-25n\\
\end{tabular}
\end{center}

\subsection{Polynomial time algorithm of factorization}
\label{factorization}
We are now able to factorize an integer in polynomial time.
Let $C>3$ be a given integer of $c$ bits, that is $2^{c-1}\le C< 2^c$. Let $A$ and $B$ be two unknown factors so that $C=A\times B$. Let $a$ and $b$ be the number of bits of $A$ and $B$ respectively. We have $a+b-1\le c \le a+b$. If $B\le A$, the trivial solution is $A=C$ and $B=1$, i.e., $a=c$ and $b=1$. All non trivial solutions with $B\le A$ are such that $a<c$ and $b \le c/2$. In the  multiplication environment, we can choose $n=c-1>1$ and $m=[(c+1)/2]>1$, given that we may complete the sets of bits by a number of zeros if necessary.
 Since we have $m+n>c $, it is convenient to define the array $c_j=0$ for $j\in\llbracket c,m+n-1\rrbracket$.
Now, we construct the LP system of the multiplication environment.
The system of both structural (Tables~\ref{equamult}, \ref{equaproduit}) and universal equations is complemented with the $m+n$ following \emph{specific data equations}:
\begin{align*}
\Prob(\mathsf{C}_j = c_j|\Pi)&=1 
\end{align*}
where $j\in\llbracket 0,m+n-1\rrbracket$.
These data equations are translated in Table~\ref{equaC} in term of global variables $\mathsf{X}_k$ (using Table~\ref{varglobmult}). 
\begin{table}
\footnotesize{
\fbox{
\begin{minipage}{\textwidth}
\begin{align*}
\Prob(1) &=  c_0 \\
\Prob(3nj-n+1) &= c_j \mathrm{~for~}  j\in\llbracket 1,m-1\rrbracket\\
\Prob(3nm-4n-m+2+j) &= c_j \mathrm{~for~}  j\in\llbracket m,m+n-2\rrbracket\\
\Prob(3nm-2n) &= c_{m+n-1} \\
\end{align*}
\end{minipage}
}
\caption{\footnotesize
\label{equaC}
Data specific equations for factorization. The numbers $c_j$ are the coefficients of the binary expansion of the input integer $C$ for $j< c$ and  $0$ for $j\ge c$. These $n+m$ equations have to be added to the set of structural equations (Tables~\ref{equamult}, \ref{equaproduit}) and to the universal equations.}
}
\end{table}

We have a LP problem of $30mn-14m-22$ unknowns and $34mn-18m-26n$ equations. 
Note that the equations have a maximum of 3 coefficients with an average of 2 entries. Therefore, the matrix is widely sparse.
Since $m=$ O$(c)$ and $n=$ O$(c)$ the dimension of the problem is O$(c^2)$. The detail is the following:
\begin{center}
\begin{tabular}{lL} 
unknowns: & 30mn-14m-22n\\
including&\\
1 literal: &2\times(3mn+m-n)\\
2 literals: &4\times (4mn-2m-3n)\\
3 literals: &8\times (mn-m-n)\\
\hline
equations: &34mn-18m-26n\\
including&\\
structural: &  3mn          -2n\\
universal: &31mn-19m-25n\\
data: &            m + n\\
\end{tabular}
\end{center}
It is suitable to simplify at once the system by use of Eq.(\ref{probaRule}).

- If the system is feasible, we have to check whether deterministic solutions exist by optimization. Generally, this can be obtained by optimazing a maximum of $2m$ objective functions. Let $k_0, k_1, \dots, k_{m-1}$ be the global labels corresponding to the variables $\mathsf{B}_0, \mathsf{B}_1\dots\mathsf{B}_{m-1}$ respectively (Table \ref{varfactors}). 
We first select the two objective functions $z_0=\Prob(\pm k_0)$. If the two maxima are both different from $1$, then $C$ is prime. Otherwise, we have a maximum $z_1=1$ for say $z_0=\Prob(-k_0)$. Now, we select two objective functions  $z_1=\Prob(-k_0) + \Prob(\pm k_1)$.  Now, if the two maxima are both different from  $2$, then again $C$ is prime. Otherwise the maximum $2$ is obtained e.g., for $z_1=\Prob(-k_0) + \Prob( k_1)$.  We iterate with $z_2=\Prob(-k_0) + \Prob( k_1)+\Prob(\pm k_2)$, etc. Finally, we select an integer $B$ with $\Prob(\pm k_i) = 1$ for $i\in\llbracket 0,m-1\rrbracket$ or otherwise $C$ is prime. Finally, we check $C/B$. Thus,  we obtain two factors $A$ and $B$ or prove that $C$ is prime in polynomial time.

- If the system is not feasible, $C$ is prime with certainty.  Again, this result is obtained in polynomial time. (However, from our computations, the system seems always feasible).

\emph{Example 1: A trivial problem.} Let $C=6$ and thus $c=3$ bits. We can choose $m=2$ and $n=2$. Even if the computation is straightforward, we have nevertheless $48$ unknowns and $48$ equations and therefore the computation is difficult to perform by hand! After Gauss elimination, the rank is $42$. The LP problem is feasible and since $m=n$ the factors $A$ and $B$ are not implicitly ordered. We obtain the expected conventional deterministic solutions $(A =2 ;\  B=3)$ and $(A = 3 ;\ B=2)$  but also a continuous set of non deterministic solutions which can be regarded as a superposition of the deterministic solutions (in the quantum sense).  
 
\emph{Example 2: A toy model for quantum mechanics.}  Let $C=5$ and thus $c=3$ bits. We can choose $m=2$ and $n=2$. We have still $48$ unknowns, $48$ equations and the rank is $42$. Again, the LP problem is feasible but only accept \emph{non-fully deterministic solution}. For instance, if we have $\Prob(\mathsf{A}_0 =1|\Pi) =1$ and $\Prob(\mathsf{B}_0 =1|\Pi) =1$, we do have  $\Prob(\mathsf{A}_1 =1|\Pi) = 1/2$ and $\Prob(\mathsf{B}_1 =1|\Pi) =1/2$. Let us open a parenthesis: When contemplating the LP problem, it resembles a quantum system. We give a few hints even if a comprehensive discussion is clearly beyond the scope of this paper: The prime `$5$' is a mathematical object defined by the variables $\mathsf{C}_j$. We decide to describe this object by a number of artificial variables, like $\mathsf{A}_i$ and $\mathsf{B}_i$. As a result, the outcomes of these new variables may be only defined in probability, depending upon the solution of the LP system. Each solution is similar to the setting of a quantum object. When a particular solution/setting is chosen, the probabilities are \emph{ipso facto} determined. They rely on the structure of internal parameters, described, e.g., by the symmetry of the feasible LP-polytope~\cite{murty}. In order to force a deterministic outcome, we may proceed to a random trial but then, the system collapses, e.g., into two integers and the original prime is destroyed. In this respect the LP system of a composite integer like `$6$' resembles a classical object with some quantum features like possible superposition of states. In our opinion, this analogy supports the conjecture that quantum formalism is a particular codification of states of knowledge, exactly as the Bayesian formulation is. We close this parenthesis.

For small values of $C$,  when compared with conventional algorithms, one can quite rightly argue that the computation complexity is out of all proportions, up to say 100 bits. Actually, the interest of the method only arises for big integers, when the conventional algorithms crash into the exponential wall, while the Bayesian route remains polynomial. The next examples are potentially in this range but are not in the present capability of this author due at least to lack of familiarity with large systems.

\emph{Example 3: The present state of the art}. Let $C$ be a $768$-bit integer. We can choose $m=384$ and $n=767$. We obtain a rough LP system with 8~813~590 unknowns and  9~987~098 equations which uses sparse matrices with three or less non-zero entries per row. The number of 768 bits corresponds to the last factorized integer of the now obsolete \emph{RSA Challenge}~\cite{RSA}. The computation was carried out in 2009 by an international team~\cite{kleinjung}, using  the best known algorithm, namely, the \emph{Number Field Sieve} (NFS)~\cite{lenstra}. According to the authors, the overall effort represent 2000 years on a single core 2.2 GHz  processor. Most of the computation consists in sieving a large number of smooth roots modulo $C$ from a pair of convenient polynomials. The last operation is the merging step for Gauss elimination.
It produced a 192~796~550 $\times$ 192~795~550 sparse matrix  with on average 144 non-zero entries per row solved in one day. Clearly, the present Bayesian method is in principle by far simplest than this last operation.

\emph{Example 4: A challenging computation beyond the present state of the art. } Let $C$ be a 1024-bit integer. We can choose $m=512$ and $n=1023$. We obtain a rough LP sparse system with 15~683~606 unknowns and  17~772~570 equations still with three or less non-zero entries per row. These dimensions may  be reduced if we know the order of magnitude of the factors. The computation is only twice more difficult than for  $768$ bits. Using the NFS algorithm, the factorization of the 1024-bit integer of the RSA challenge is presently out of reach but expected by the year 2020 .

\emph{Exemple 5: An outstanding challenging computation.} Let $C$ be a $2048$-bit integer. The rough system dimensions will be about  $63.~10^6\times 71.~10^6$. This factorization is definitively out of reach of the NFS algorithm but, in principle, still tractable with the present Bayesian method.

\section{Conclusion}
The paper applies previous results on the conversion of an ensemble of Boolean formulae into a linear programming problem: Now, we introduce the concept of `Bayesian arithmetic'. In this model, a Diophantine equation is interpreted as a set of prior conditions in the framework of Bayesian probability theory.  We have shown that the dimension of the LP problem scales as O$(n)$ for the sum of two $n$-bit integers and as O$(mn)$ for the product of two integers of $m$ and $n$ bits respectively. As a result, the LP problem encoding a Diophantine equation is solved in polynomial time in the number of working bits. 

The first significant application is a polynomial time algorithm of factorization.  This stresses the need to revisit all current public key encryption systems.

A similar approach should solve in polynomial time any logical function defined by a set of truth tables. 

Finally, the exponential speed-up over conventional algorithms recalls quantum computation. This questions about a deep similarity between LP and quantum mechanics formalism.


\bibliography{biblio}

\end{document}